\documentclass[showkeys,showpacs]{revtex4}

\usepackage{graphicx}
\usepackage{amsmath}
\usepackage{amssymb}
\usepackage{color}

\begin{document}

\title{Long-term bounce of the Universe filled with the Higgs field}

\author{
Vladimir Dzhunushaliev
\footnote{
Email: vdzhunus@krsu.edu.kg}
}
\affiliation{Department of Physics and Microelectronic
Engineering, Kyrgyz-Russian Slavic University, Bishkek, Kievskaya Str.
44, 720021, Kyrgyz Republic \\
and \\
Institute of Physicotechnical Problems and Material Science of the NAS
of theKyrgyz Republic, 265 a, Chui Street, Bishkek, 720071,  Kyrgyz Republic
}

\author{Kairat Myrzakulov}
\affiliation{Dept. Gen. Theor. Phys., Eurasian National University, Astana, 010008, Kazakhstan}

\author{Ratbay Myrzakulov}
\affiliation{Dept. Gen. Theor. Phys., Eurasian National University, Astana, 010008, Kazakhstan
}
\email{cnlpmyra1954@yahoo.com}

\begin{abstract}
The cosmological solution with a long-term bouncing off time is presented. The solution has a preceding contracting and a subsequent expanding phases and between them there exists a bouncing off phase with arbitrary time duration.
\end{abstract}

\maketitle

\section{Introduction}

The standard cosmological model (for review see \cite{pdg}) gives us an accurate description of the evolution of the universe. In spite of its success, the standard cosmological model has a series of problems such as the initial singularity, the cosmological horizon, the flatness problem, the baryon asymmetry, and the nature of dark matter and dark energy.

Under the dynamical laws of general relativity, the standard FLRW cosmology becomes singular at the origin of Universe. The matter density and geometrical invariants diverge as the volume of the Universe goes to zero. The Big Bang singularity seems to be an unavoidable aspect of the currently established cosmological model \cite{Weinberg08} which probably only a full quantum theory of gravity could resolve. A bouncing universe with an initial contraction to a non-vanishing minimal radius, then subsequent an expanding phase provides a possible solution to the singularity problem of the standard Big-Bang cosmology.

Bouncing cosmologies, in which the present era of expansion is preceded by a contracting phase, have been studied as potential alternatives to inflation in solving the problems of standard FRW cosmology. The first explicit semi-analytic solution for a closed bouncing FRW model filled 
by a massive scalar field was found by Starobinskii~\cite{starobinskii}. Later explicit solutions for a bouncing geometry were obtained by Novello and Salim \cite{NovelloSalim}, and Melnikov and Orlov \cite{melni}. For the review of the cosmological bounce one can see review \cite{Novello:2008ra}.

In this paper we would like to show that a Universe bounce can be not only a short time event but it can be a long-term process.

\section{The statement of the problem}

In this section we will investigate a cosmological solution for a closed Universe filled with a Higgs scalar field. Our goal is to find a regular solution with contracting, expanding and bouncing off phases.

We start with the Lagrangian
\begin{equation}
\label{lagrangian}
  L =-\frac{R}{2}+
	\varkappa \left[
		\frac{1}{2}\partial_\mu \phi \partial^\mu\phi - V(\phi)
		\right]
\end{equation}
where $R$ is the 4D scalar curvature, $\phi$ is the scalar Higgs field with the potentials  $V(\phi)$. The corresponding field equations are
\begin{eqnarray}
	R_{\mu \nu} - \frac{1}{2} g_{\mu \nu} R &=& \varkappa T_{\mu \nu},
\label{2-10}\\
	\frac{1}{\sqrt{-g}} \frac{\partial }{\partial x^\mu}
	\left(
		\sqrt{-g} g^{\mu \nu} \frac{\partial \phi}{\partial x^\nu}
	\right) &=& - \frac{d V(\phi)}{d \phi}.
\label{2-20}
\end{eqnarray}
the potential $V(\phi)$ is the Mexican hat potential
\begin{equation}
	V(\phi) = \frac{\lambda}{4} \left(
		\phi^2 - m^2
	\right)^2 - V_0
\label{2-30}
\end{equation}
where $m$ and $V_0$ are constants and $V_0$ can be considered as a cosmological constant. The energy-momentum tensor $T_{\mu \nu}$ for the scalar field is
\begin{equation}
	T_{\mu \nu} = \left( \nabla_\mu \phi \right)
	\left( \nabla_\nu \phi \right) - g_{\mu \nu} L.
\label{2-50}
\end{equation}
For the investigation of bouncing off of Universe we consider the cosmological metric
\begin{equation}
	ds^2 = dt^2 - a^2(t) \left[
		d \chi^2 + \sin^2 \chi \left(
			d \theta ^2 + \sin^2 \theta d \varphi^2
		\right)
	\right] .
\label{2-60}
\end{equation}
After the substitution of the metric \eqref{2-60} into field equations \eqref{2-10} \eqref{2-20} we have following equations set
\begin{eqnarray}
	\frac{3 \dot a^2}{a^2} + \frac{3}{a^2} &=& \varkappa
	\left[
		\frac{\dot \phi^2}{2} + \frac{\lambda}{4}
		\left( \phi^2 - m^2 \right)^2 - V_0	
	\right] ,
\label{2-70}\\
  \frac{2 \ddot a}{a} + \frac{\dot a^2}{a^2} + \frac{1}{a^2} &=& \varkappa
	\left[
		- \frac{\dot \phi^2}{2} + \frac{\lambda}{4}
		\left( \phi^2 - m^2 \right)^2 - V_0	
	\right] ,
\label{2-80}\\
  \ddot \phi + \frac{3 \dot a \dot \phi}{a} &=& \lambda \phi
  \left( m^2 - \phi^2 \right).
\label{2-90}
\end{eqnarray}
To begin with we would like to define the notion ``bouncing off of the Universe''. We will say, that the Universe undergo bouncing off if enough a long compression stage is replaced by enough a long stage of the expansion. A long time interval $\Delta t$ means that $\Delta t \gg t_{Planck}$ or $\Delta t \approx t_{Universe}$ where $t_{Universe}$ is the life time of the Universe.

Why we need for such definition ? Below we will show that bouncing off can be a complicated process lasting a long enough time.

The statement of the problem: \textcolor{blue}{\emph{to find solutions having big enough compression/expansion stages and between them a long enough bouncing off time interval.}}

\section{A long time bouncing off solutions}

\begin{figure}[h]
\begin{minipage}[t]{.45\linewidth}
 \begin{center}
 \fbox{
  	\includegraphics[height=.8\linewidth,width=.8\linewidth]{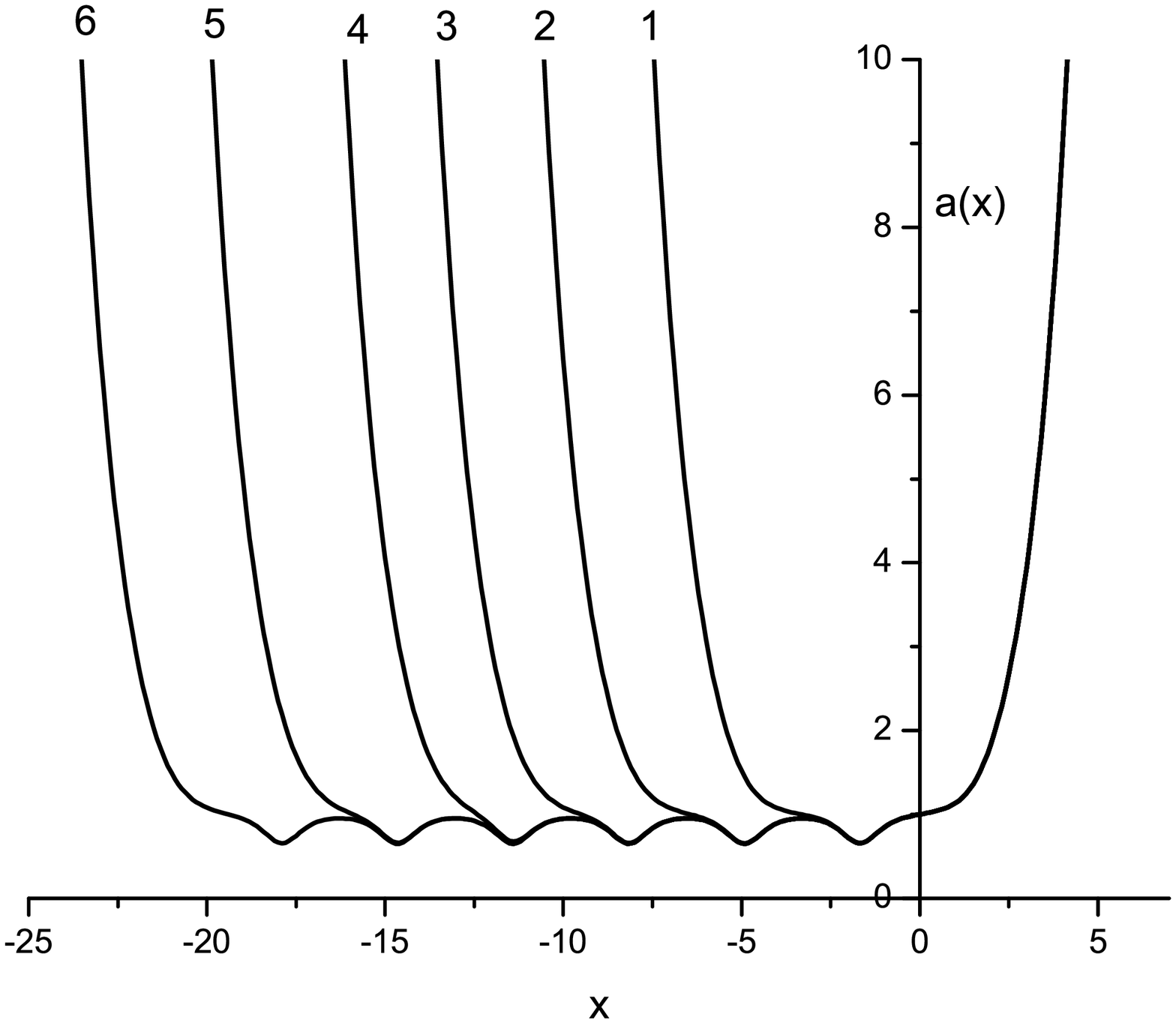}
  }
  \caption{The profiles $a(t)$. The curve 1 for $a_1 = 0.08$,
  the curve 2 for $a_1 = 0.0805$,
  the curve 3 for $a_1 = 0.080502$,
  the curve 4 for $a_1 = 0.0805020498$,
  the curve 5 for $a_1 = 0.080502049893$,
  the curve 6 for $a_1 = 0.08050204989312597$.}
  \label{fg1}
	\end{center}
\end{minipage}\hfill
\begin{minipage}[t]{.45\linewidth}
 \begin{center}
 \fbox{
	  \includegraphics[height=.8\linewidth,width=.8\linewidth]{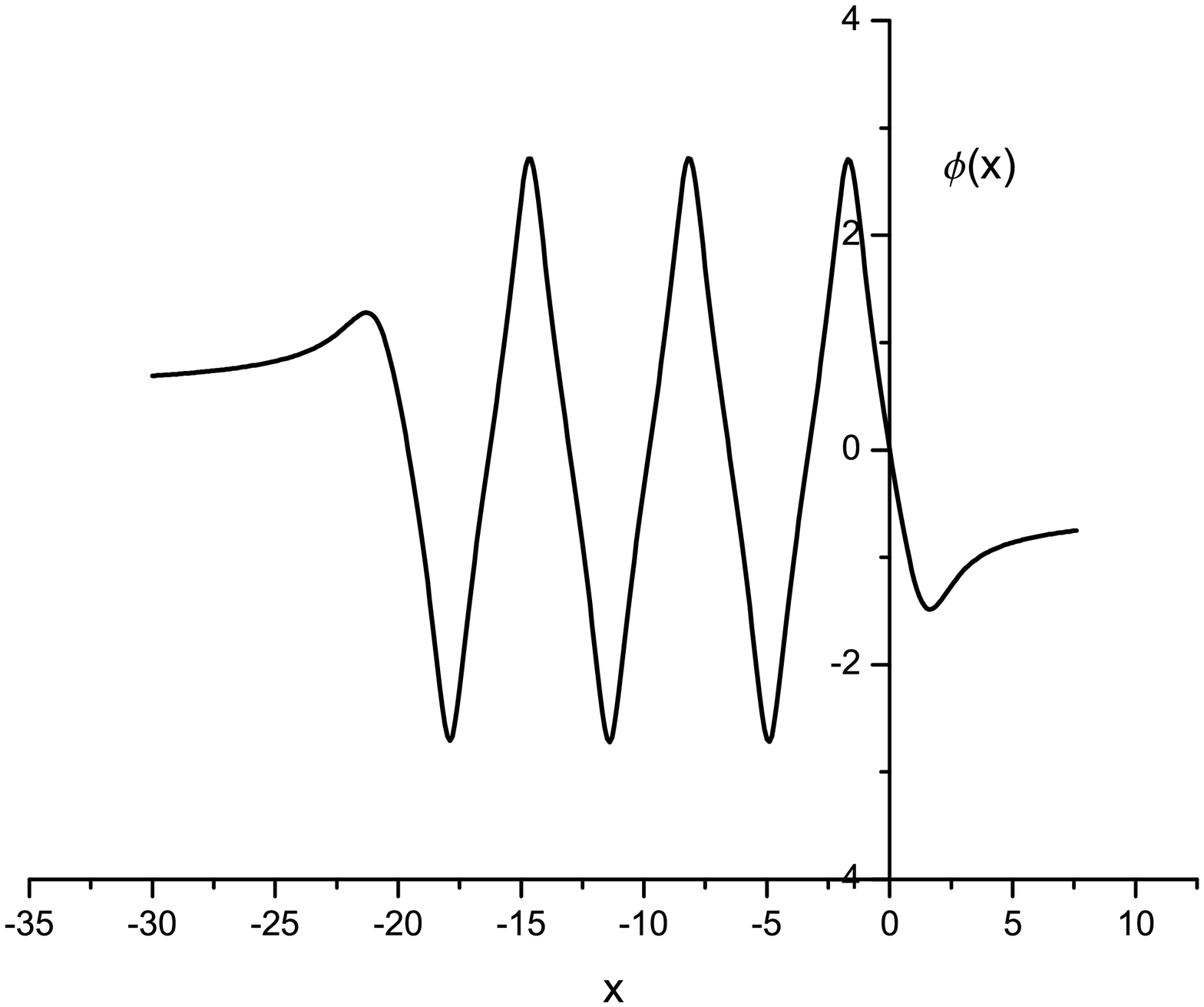}
 }
	\caption{The profile $\phi(t)$ for $a_1 = 0.08050204989312597$.}
	\label{fg2}
	\end{center}
\end{minipage}\hfill
\end{figure}

\begin{figure}[h]
\begin{minipage}[t]{.45\linewidth}
 \begin{center}
 \fbox{
  	\includegraphics[height=.8\linewidth,width=.8\linewidth]{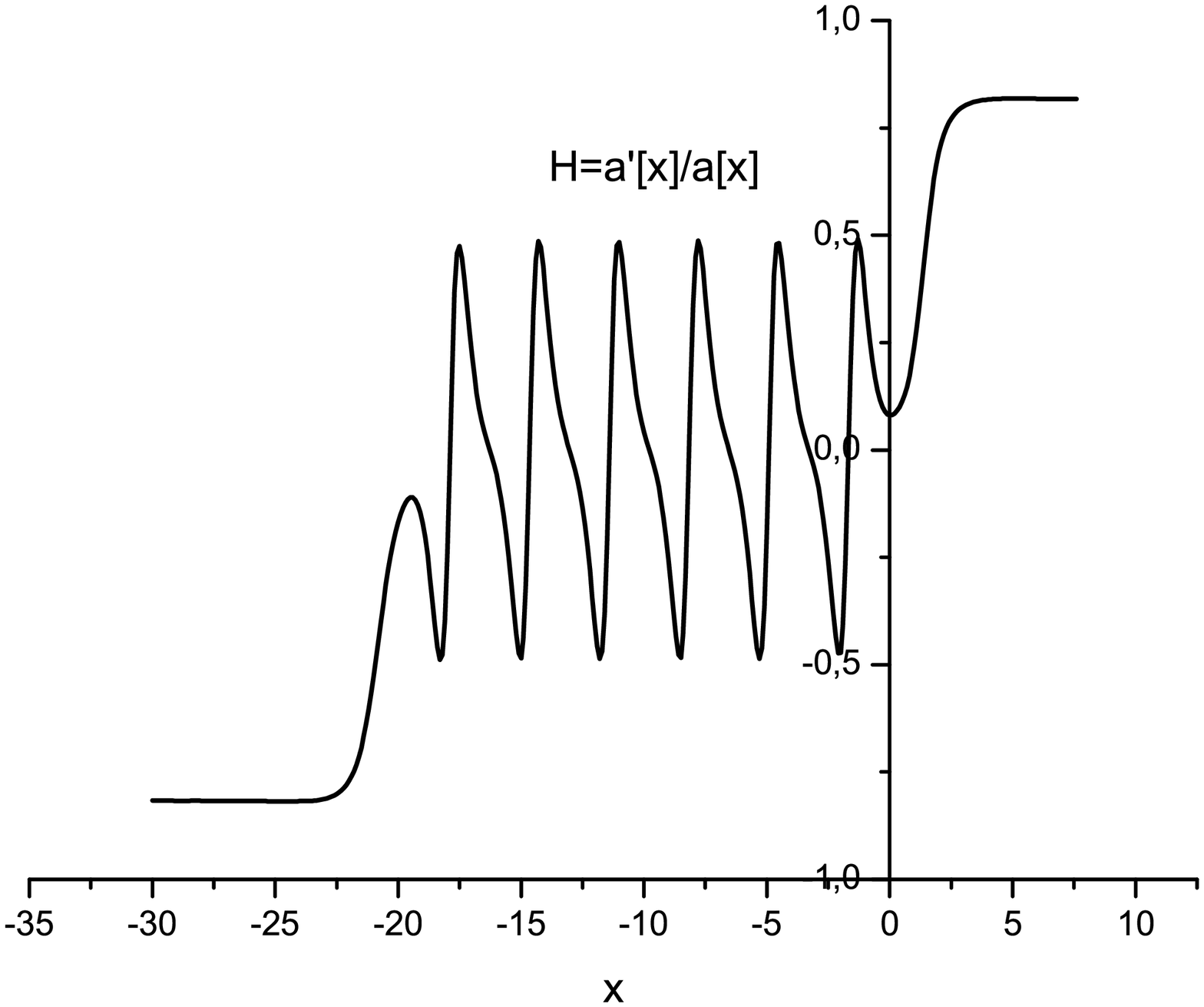}
  }
  \caption{The profile of Hubble parameter $H(t) = \frac{\dot a(t)}{a(t)}$
	for $a_1 = 0.08050204989312597$.}
  \label{fg3}
	\end{center}
\end{minipage}\hfill
\begin{minipage}[t]{.45\linewidth}
 \begin{center}
 \fbox{
	  \includegraphics[height=.8\linewidth,width=.8\linewidth]{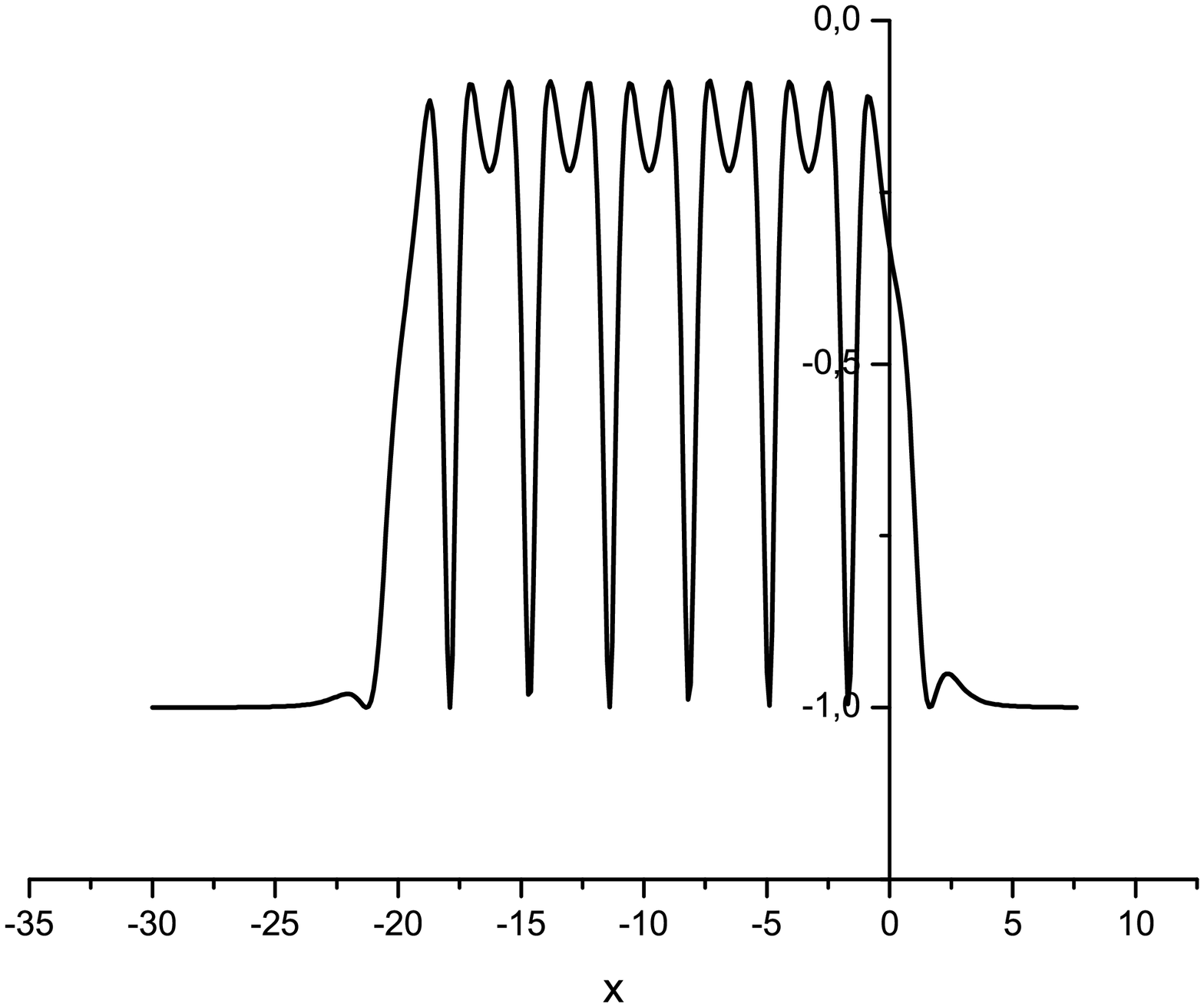}
 }
	\caption{The profile of the equation state $w(t)$ for $a_1 = 0.08050204989312597$.}
	\label{fg4}
	\end{center}
\end{minipage}\hfill
\end{figure}

\begin{figure}[h]
\begin{minipage}[t]{.45\linewidth}
 \begin{center}
 \fbox{
  	\includegraphics[height=.8\linewidth,width=.8\linewidth]{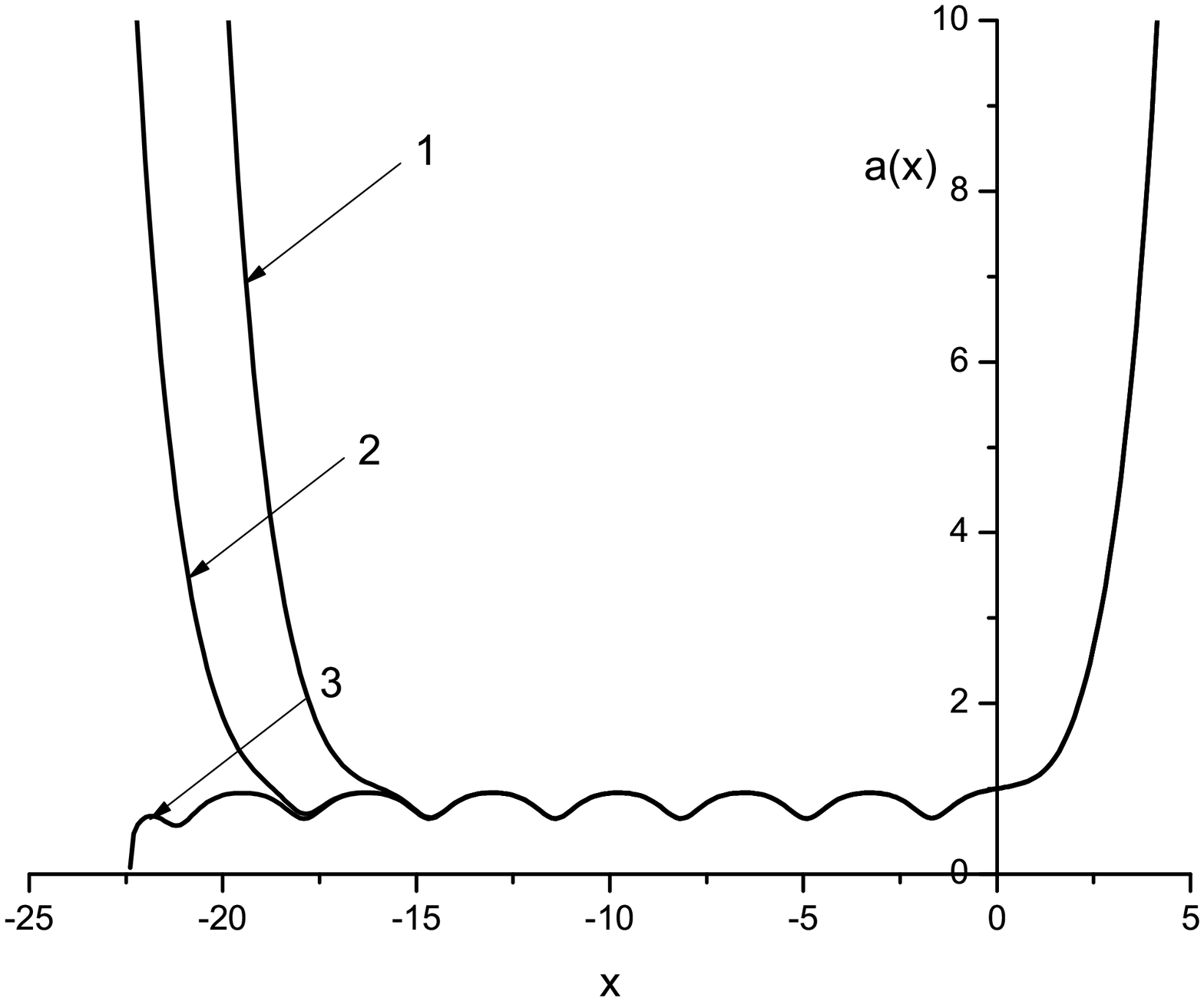}
  }
  \caption{The profiles of regular and singular $a(t)$. The regular solutions:
  curve 1 for $a_1 = 0.080502049893$,
  curve 2 for $a_1 = 0.08050204989312597$. The singular solutions:
  curve 3 for $a_1 = 0.08050204989312599$.}
  \label{fg5}
	\end{center}
\end{minipage}\hfill
\end{figure}

We will investigate a long term bouncing off solutions with a point of inflection with following conditions
\begin{equation}
	\ddot a(t) \begin{cases}
		< 0, 			& \text{if } t < t_0, \\
		= 0, 			& \text{if } t = t_0, \\
		> 0, 			& \text{if } t > t_0
	\end{cases}
\label{3-10}
\end{equation}
here $t_0$ is the point of inflection. In the consequence of the existence of the inflection point the functions $a(t), \phi(t)$ can be presented in the form
\begin{eqnarray}
	a(t) &=& a_0 + a_1 (t-t_0) + a_3 (t-t_0)^3 + \cdots ,
\label{3-20}\\
  \phi(t) &=& \phi_0 + \phi_1 (t-t_0) + \phi_2 (t-t_0)^2 + \phi_3 (t-t_0)^3 + \cdots .
\label{3-30}
\end{eqnarray}
Using the conditions \eqref{3-10} and equations \eqref{2-70} - \eqref{2-90} one can find the following constraints on the initial conditions
$a_0 = a(t_0), a_1 = \dot a_0 = \dot a(t_0), \phi_0 = \phi(0), \phi_1 = \dot \phi_0 = \dot \phi(0)$ and the cosmological constant $V_0$
\begin{eqnarray}
	\frac{2 \dot a^2_0}{a^2_0} + \frac{2}{a^2_0} &=& \varkappa
	\left[
		\frac{\lambda}{4} \left( \phi^2_0 - m^2 \right)^2 - V_0	
	\right] ,
\label{2-110}\\
  \frac{2 \dot a^2_0}{a^2_0} + \frac{2}{a^2_0} &=& \varkappa
	\dot \phi^2_0 .
\label{2-120}
\end{eqnarray}
It means that the cosmological constant $V_0$ is defined uniquely
\begin{equation}
	\Lambda = \varkappa V_0 =  \frac{\lambda}{4} \varkappa \left( \phi^2_0 - m^2 \right)^2 -
	2 \left(
		\frac{\dot a_0^2}{a_0^2} + \frac{1}{a_0^2}.
	\right)
\label{2-130}
\end{equation}
For the numerical investigation we introduce the dimensionless time
$x = t/\sqrt{\varkappa}$ and dimensionless functions
$\phi \sqrt{\varkappa} \rightarrow \phi$, $a/\sqrt{\varkappa} \rightarrow a$. Then equations \eqref{2-70}-\eqref{2-90} becomes
\begin{eqnarray}
	\frac{3 \dot a^2}{a^2} + \frac{3}{a^2} &=& \left[
		\frac{\dot \phi^2}{2} + \frac{\lambda}{4}
		\left( \phi^2 - m^2 \right)^2 - V_0	
	\right] ,
\label{2-140}\\
  \frac{2 \ddot a}{a} + \frac{\dot a^2}{a^2} + \frac{1}{a^2} &=& \left[
		- \frac{\dot \phi^2}{2} + \frac{\lambda}{4}
		\left( \phi^2 - m^2 \right)^2 - V_0	
	\right] ,
\label{2-150}\\
  \ddot \phi + \frac{3 \dot a \dot \phi}{a} &=& \lambda \phi
  \left( m^2 - \phi^2 \right)
\label{2-160}
\end{eqnarray}
with the following initial conditions
\begin{equation}
	a(0) = a_0, \quad \dot a(0) = a_0, \quad
	\phi(0) = \phi_0, \quad
	\dot \phi_0 = - \frac{\sqrt{2 \dot a_0^2 + 2}}{a_0}.
\label{2-170}
\end{equation}
The numerical solutions are presented in Fig's \ref{fg1}-\ref{fg5}. In Fig. \ref{fg1} the profiles for different $a_1$ are presented. From this figure we see that there exist the bounce with different time duration. Every such solution can be enumerated with the number of minima or maxima. On Fig's \ref{fg1}-\ref{fg4} the profiles of $\phi(t)$, Hubble parameter $H(t) = \frac{\dot a(t)}{a(t)}$ and the state equation $w(t)$ for $a_1 = 0.08050204989312597$ are presented. It is useful to present the profile of a state equation (see, Fig. \ref{fg4}) in the form
\begin{equation}
	w = \frac{p}{\varepsilon} = \frac{T^{11}}{T^{00}} =
	\frac{\frac{\dot \phi^2}{2} - \frac{\lambda}{4} \left( \phi^2-m^2 \right) + V_0}
	{\frac{\dot \phi^2}{2} + \frac{\lambda}{4} \left( \phi^2-m^2 \right) - V_0}
\label{2-210}
\end{equation}
where $p$ is the pressure and $\varepsilon$ is the energy density. We see that in the contracting and expanding stages $w \approx -1$ and additionally in during of bouncing off stage there are points where $w \approx -1$.

How long could be such prolonged bouncing off ? For the investigation of this question we are addressing to Fig. \ref{fg5}. We see that for $a_1 = 0.080502049893, 0.08050204989312597$ the solution has bouncing off but for $a_1 = 0.08050204989312599$ the solution is singular one. It means that there exists a special solution with $a_1 = a_1^*$ where $0.08050204989312597 < a_1^* < 0.08050204989312599$. From Fig. \ref{fg1} and Fig. \ref{fg5} we see that such solution should have an infinite long bouncing off with a contracting phase at $t = -\infty$ and an expansion phase at $t = +\infty$. The most interesting is that by $a_1$  close enough to $a_1^*$ we may have a bouncing off time with any duration.

In during of bouncing off the size of Universe is $a \approx a_{Pl}$. It means that the existence time of Universe in such state can be somehow long with possible expansion at any time.

\section{Discussion and conclusions}

Here we have shown that the process of Universe bouncing off is not a simple process. Implicitly is supposed that bouncing off happens at \textcolor{blue}{\emph{a moment}}. Here we have shown that it can be \textcolor{blue}{\emph{a long-term process}} taking place at the Planck region. It allows us to consider a quantum birth of the Universe as following process: the Universe exists in the state with
$a_1 = a_1^*$ and in the consequence of quantum fluctuations of $a_1$ the Universe pass to the state with $a_1 \approx a_1^*$. After that the Universe go into an inflation phase.

\end{document}